\documentclass[aps,pra,eqsecnum,twocolumn]{revtex4}
\usepackage{graphicx}
\usepackage{bm}
\begin{document}
\preprint{}
\title{Practical schemes for the measurement of
angular-momentum covariance matrices in quantum optics}
\author{\'{A}ngel Rivas}
\email{A.Rivas@herts.ac.uk}
\affiliation{School of Physics, Astronomy and Mathematics,
University of Hertfordshire College Lane, Hatfield, Hertfordshire,
AL10 9AB, United Kingdom}
\author{Alfredo Luis}
\email{alluis@fis.ucm.es}
\homepage{http://www.ucm.es/info/gioq}
\affiliation{Departamento de \'{O}ptica, Facultad de Ciencias
F\'{\i}sicas, Universidad Complutense, 28040 Madrid, Spain }
\date{\today}

\begin{abstract}
We develop practical schemes for the measurement of the
covariance matrix for intrinsic angular-momentum variables
in quantum optics. We particularize this approach to
two-beam polarimetry and interferometry, as well as to
ensembles of two-level atoms interacting with classical
fields. We show the practical advantages of noisy simultaneous
measurements.
\end{abstract}


\maketitle

\section{Introduction}

Angular-momentum variables represent basic observables both in
classical and quantum optics, especially in three fundamental
areas: polarization, interferometry, and light-matter interaction
\cite{HL,be,yea,mbe,cvpe,asc,yop,SZ,yobs,pepe}. For definiteness,
throughout we focus on intrinsic (not orbital) angular momenta.
This is for example the case of the Stokes parameters, which
provide a complete account of second-order (in complex amplitudes)
statistical properties of two-mode polarization and interference.
Moreover, spin operators are basic in atomic physics such as
in the case of ensembles of two-level atoms described individually
as spin 1/2 systems.

The second-order statistics of angular-momentum variables are
crucial in diverse areas. This is the case with quantum metrology,
where angular-momentum statistics determine the ultimate limit to
the resolution of interferometric and spectroscopic measurements
\cite{HL,be,yea}. Moreover, angular-momentum covariance matrices
enter in the analysis of many-body entanglement \cite{mbe}, in
continuous-variable polarization entanglement \cite{cvpe}, and
for light-mediated detection of atomic-spin correlations
\cite{asc}.

Recently we have proposed an SU(2)-invariant characterization
of angular-momentum fluctuations via the diagonalization
of the covariance matrix \cite{RL07}. Invariance under SU(2)
transformations is a desirable property since two states
connected by a deterministic SU(2) transformation should be
statistically equivalent. Similar invariance ideas are at the
heart of current investigations about the coherence between
classical vectorial waves \cite{cvw}.

In this work we develop simple practical schemes to determine
experimentally the angular-momentum covariance matrix of a
given system in an unknown state. We particularize the method
to diverse optical two-mode polarimetric and interferometric
configurations, as well as to ensembles of two-level atoms.
It is worth stressing that this analysis applies equally well
to quantum and classical optics. In the classical domain the
situation is much more simple since in principle one can always
perform as many simultaneous measurements as desired of any set
of observables in accurate copies of the original beam provided
by beam splitting, for instance. This idea can be fruitfully
translated to the quantum domain in the form of noisy simultaneous
measurements of noncommuting angular-momentum components.

In Sec. II we recall basic definitions and results. In Sec. III
we present a basic scheme for the measurement of angular-momentum
covariance matrices which is particularized to polarimetric,
interferometric, and spectroscopic situations. In Sec. IV we
present an interferometric noisy simultaneous measurement of
angular-momentum components providing a simple and exact
practical determination of the covariance matrix with a single
experimental configuration. In Sec. V we consider the bright
limit in which the angular-momentum covariance matrix becomes a
quadrature (or position-linear momentum) covariance matrix.

\section{Definitions}

\subsection{Definition and two-mode realization}

Let us consider arbitrary dimensionless angular momentum
operators $\bm{j}^t = (j_1, j_2 , j_3 )$, where the
superscript $t$ denotes matrix transposition. In quantum
physics they are defined by the fulfillment of the commutation
relations
\begin{equation}
\label{cr}
[ j_k ,j_\ell ] = i \sum_{n=1}^3 \epsilon_{k,\ell ,n}
j_n ,
\qquad [j_0,\bm{j} ] = \bm{0} ,
\end{equation}
where $\epsilon_{k,\ell,n}$ is the fully antisymmetric
tensor with $\epsilon_{1,2,3} =1$ and $j_0$ is defined
by the relation
\begin{equation}
\label{j2}
\bm{j}^2 = j_0 \left ( j_0 + 1 \right ) .
\end{equation}
For the sake of completeness we take into account that
$j_0$ may be an operator. This is the case of two-mode
realizations where $j_0$ is proportional to the number
of photons. In classical optics the situation is
similar by replacing commutators by Poisson brackets and
Eq. (\ref{j2}) by $\bm{j}^2 = j^2_0$ (with $\langle
\bm{j} \rangle^2 \leq \langle j_0 \rangle^2$, where the
brackets denote an ensemble average).

In quantum and classical optics two-mode realizations of
angular momentum play a relevant position in polarimetry and
interferometry. In the quantum case, denoting by $a_{1,2}$
the complex amplitudes operators of two field modes, we get
that
\begin{eqnarray}
\label{So}
j_0 = \frac{1}{2} \left ( a^\dagger_1 a_1 + a^\dagger_2
a_2 \right ), & &
j_1 = \frac{1}{2} \left ( a^\dagger_2 a_1 + a^\dagger_1
a_2 \right ) , \nonumber \\ & & \\
j_2 = \frac{i}{2} \left ( a^\dagger_2 a_1 - a^\dagger_1
a_2 \right ),  & &
j_3 = \frac{1}{2} \left ( a^\dagger_1 a_1 - a^\dagger_2
a_2 \right ) , \nonumber
\end{eqnarray}
satisfy Eqs. (\ref{cr}) and (\ref{j2}) \cite{SCH}, where the
superscript $\dagger$ denotes Hermitian conjugation. In the
classical domain $a_{1,2}$ are classical amplitudes so that
Hermitian conjugation $a^\dagger_{1,2}$ is replaced by complex
conjugation $a^\ast_{1,2}$.

In polarimetry these are essentially the Stokes variables.
The normalized vector $\langle \bm{j} \rangle /\langle j_0
\rangle$ defines the Poincar\'{e} sphere as a suitable
representation of polarization states and transformations
\cite{cvpe,pepe}. These are also basic variables in two-beam
interferometry. For example, the third component $j_3$ is
proportional to the difference of the number of photons between
two modes, while $j_{1,2}$ express the coherence between the
interfering beams.

\subsection{Covariance matrix}

The complete second-order statistics of $\bm{j}$ is contained
in the $3 \times 3$ real symmetric covariance matrix $M$
\begin{equation}
\label{Mme}
M_{k, \ell} = \frac{1}{2} \left ( \left \langle
j_k j_\ell \right \rangle + \left \langle j_\ell
j_k \right \rangle \right ) - \left \langle j_k
\right \rangle \left \langle j_\ell  \right \rangle ,
\end{equation}
with $M^t = M$ and $M^\ast = M$. The alternative definition
$M^\prime_{k, \ell} = \left \langle j_k j_\ell \right \rangle
- \left \langle j_k \right \rangle \left \langle j_\ell
\right \rangle$ is identical to $M$ in the classical case,
while in the quantum domain it provides a complex Hermitian
matrix that contains essentially the same information as
$M$ \cite{RL07}.

The covariance matrix $M$ allows us to compute the variance
$( \Delta j_u )^2$ of an arbitrary angular-momentum component
$j_u = \bm{u} \cdot \bm{j}$, where $\bm{u}$ is any unit real
vector,
\begin{equation}
\label{utMu}
 \left ( \Delta j_u \right )^2 = \bm{u}^t M \bm{u} ,
\end{equation}
as well as the symmetric correlation of two arbitrary
components $j_u = \bm{u} \cdot \bm{j}$ and $j_v = \bm{v}
\cdot \bm{j}$, where $\bm{u}$ and $\bm{v}$ are unit real
vectors,
\begin{equation}
\frac{1}{2} \left ( \langle j_u j_v \rangle +
\langle j_v j_u \rangle \right ) -
\langle j_u \rangle \langle j_v \rangle =
\bm{v}^t M \bm{u} = \bm{u}^t M \bm{v}.
\end{equation}
Since $M$ is real and symmetric, the transformation that
renders $M$ diagonal is a rotation matrix $R_d$
\begin{equation}
M = R^t_d \pmatrix{ ( \Delta J_1 )^2 & 0 & 0 \cr
 0 & ( \Delta J_2 )^2 & 0 \cr
0 & 0 & ( \Delta J_3 )^2 } R_d .
\end{equation}
The eigenvalues of $M$, $( \Delta J_k )^2$, $k=1,2,3$, are
the variances of the components $J_k = \bm{u}_k \cdot \bm{j}$,
where $\bm{u}_k$ are the three real orthonormal eigenvectors
of $M$
\begin{equation}
\label{MD}
M \bm{u}_k =  ( \Delta J_k )^2  \bm{u}_k .
\end{equation}
Following standard nomenclature in statistics we refer to
$\bm{J}$ and $\Delta \bm{J}$ as principal components and
principal variances, respectively. We stress that both
$\bm{J}$ and $\Delta \bm{J}$ depend on the system state.
The principal variances provide an SU(2) invariant
characterization of angular momentum fluctuations
\cite{RL07}.

\subsection{SU(2) invariance}

Throughout, by SU(2) invariance we mean the statistical
equivalence between states connected by unitary
deterministic transformation generated by $\bm{j}$
\begin{equation}
\label{U}
U = \exp \left ( i \theta \bm{u} \cdot \bm{j} \right )  ,
\end{equation}
where $\theta$ is a real parameter and $\bm{u}$ is a
unit three-dimensional real vector. It can be seen
(for example, by using the $j_0 = 1/2$ representation)
that the action of $U$ on $\bm{j}$ is a rotation $R$
of angle $\theta$ and axis $\bm{u}$ \cite{CS}
\begin{equation}
\label{rot}
U^\dagger \bm{j} U = R \bm{j} ,
\quad
U^\dagger j_0 U = j_0 ,
\end{equation}
where the $3 \times 3$ real matrix $R$ is
\begin{equation}
\label{Rkl}
R_{k,\ell} = \frac{1}{2} \textrm{tr} \left ( \sigma_\ell
\mathcal{U}^\dagger \sigma_k \mathcal{U} \right ) ,
\end{equation}
with
\begin{equation}
\label{cU}
\mathcal{U}= \exp \left ( i \theta V \right ) ,
\quad
V = \frac{1}{2} \sum_{k=1}^3 u_k \sigma_k ,
\end{equation}
where $k,\ell=1,2,3$, $\sigma_{1,2,3}$ are the Pauli
matrices and it holds that $R^t = R^{-1}$, and
$\mathcal{U}^\dagger = \mathcal{U}^{-1}$.

The SU(2) invariance of principal variances holds because
under any SU(2) transformation $M$ transforms as $M
\rightarrow R M R^t$. Therefore, the covariance matrix
$R M R^t$ associated with the transformed state has the same
principal variances as the covariance matrix $M$
associated with the original state.

In other words, the SU(2) invariance is just the mathematical
statement corresponding to the fact that the conclusions
which one could draw from an angular momentum measurement
must be independent of which set of three orthogonal
angular momentum components one chooses.

In the case of the two-mode bosonic realizations (\ref{So})
we have
\begin{equation}
\label{io}
\bm{b} = U^\dagger \bm{a} U = \mathcal{U} \bm{a},
\quad
U =\exp \left ( i \theta \bm{a}^\dagger V \bm{a} \right ) ,
\end{equation}
where $\mathcal{U}$ and $V$ are in Eq. (\ref{cU}) and
$\bm{a}^t = (a_1 ,a_2)$ and $\bm{b}^t = (b_1, b_2)$ are the
original and transformed complex amplitudes, respectively.
In this case SU(2) transformations describe basic lossless
polarization and interference elements, such as beam splitters,
phase plates, two-beam interferometers, Faraday rotators, etc.
\cite{HL,be,yea,cvpe,yop,yobs}.

\section{Practical determination of the covariance matrix}

The complete determination of the covariance matrix in a
given basis of components $\bm{j}^t = (j_1 ,j_2, j_3)$ can
be achieved by measurement of the variances of the six
operators
\begin{eqnarray}
\label{mc}
& j_{1 \pm 2} = \frac{1}{\sqrt{2}} \left ( j_1 \pm j_2
\right ) , & \nonumber \\
& j_{1 \pm 3} = \frac{1}{\sqrt{2}} \left ( j_1 \pm j_3
\right ) , & \nonumber \\
& j_{2 \pm 3} = \frac{1}{\sqrt{2}} \left ( j_2 \pm j_3
\right ) . &
\end{eqnarray}
More specifically, since (respecting the quantum lack of
commutation)
\begin{equation}
j^2_{k \pm \ell} = \frac{1}{2} \left ( j_k^2 +
j_\ell^2 \pm j_k j_\ell \pm j_\ell j_k \right ),
\end{equation}
we get that the nondiagonal matrix elements $M_{k,\ell}$,
$k \neq \ell$,  are given by
\begin{equation}
M_{k,\ell} = \frac{1}{2} \left [ \left ( \Delta j_{k + \ell}
\right )^2 - \left ( \Delta j_{k - \ell} \right )^2 \right ] .
\end{equation}
For the diagonal elements we have
\begin{eqnarray}
M_{3,3} = &\left( \Delta j_3  \right)^2 =\frac{1}{2}
\left[ \left( \Delta j_{1 + 3} \right)^2 +
\left( \Delta j_{1 - 3} \right)^2 +
\left( \Delta j_{2 + 3} \right)^2 \right.\nonumber\\
&\left. +\left( \Delta j_{2 - 3} \right)^2 -
\left( \Delta j_{1 + 2} \right)^2 -
\left( \Delta j_{1 - 2} \right)^2 \right],
\end{eqnarray}
and similarly for $\Delta j_1$ and $\Delta j_2$ by cyclic
permutations of the indices.

Note that $M$ has just six independent components because of
reality and symmetry, which agrees with the above number of
independent measured variances. The six operators (\ref{mc})
are not, strictly speaking, independent since we have, for
example,
\begin{equation}
j_{2+3} = j_{1+2} - j_{1-3},  \quad
j_{2-3} = j_{1+2} - j_{1+3}.
\end{equation}
Nevertheless, measurement of the components $j_{2\pm 3}$
is necessary to derive all $j_k j_\ell$ correlations
exclusively in terms of variances.

When one of the components of $\bm{j}$, say $j_3$, is a
principal component, the process is much more simple since
we know in advance that all the correlations between $j_3$
and $j_1$, $j_2$ vanish. Then, only four variances are
necessary: namely $\Delta j_1$, $\Delta j_2$, $\Delta j_3$,
and $\Delta j_{1+2}$ for example.

The measured components $j_{p \pm q}$ can be related with
the original ones $j_\ell$ by simple SU(2) transformations of
the form
\begin{equation}
\label{Ukm}
U_{k,\pm m} =\exp \left ( i \theta_{\pm m} j_k \right ) ,
\end{equation}
with $\theta_{\pm m} = \pm \pi/m$ and $m=2,4$, which produce
rotations of angles $\pi/2$ and $\pi/4$ around the axis
$j_k$. This is useful because $U_{k,\pm 4}$ transform the
measurement of the components $j_\ell$ in the transformed
state into the measurement of the operators $j_{p \pm q}$
in the original state, while $U_{k,\pm 2}$ transform the
components $j_\ell$ among themselves. The proper use of
these transformations is illustrated by the following particular
cases.

\subsection{Polarimetry}

\begin{figure}
\begin{center}
\includegraphics[width=6cm]{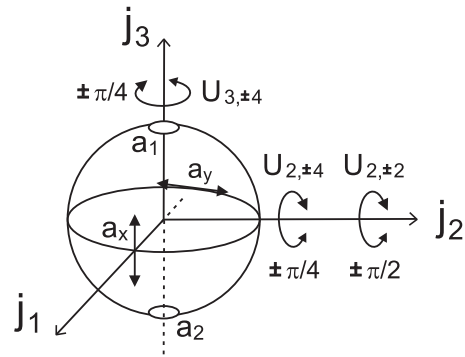}
\end{center}
\caption{Poincar\'{e} sphere illustrating polarization states
(circular at the poles and linear at the equator) and the
action of the transformations (\ref{br}).}
\end{figure}

Polarization states and transformations can be properly
represented in the Poincar\'{e} sphere, as illustrated in
Fig. 1. For definiteness we consider $a_{1,2}$ in
Eq. (\ref{So}) as the amplitudes of circularly polarized
modes,
\begin{equation}
a_1 = \frac{1}{\sqrt{2}} \left ( a_x + i a_y \right ) ,
\quad
a_2 = \frac{1}{\sqrt{2}} \left ( a_x - i a_y \right ) ,
\end{equation}
where $a_{x,y}$ are the complex amplitudes of modes linearly
polarized along the Cartesian axes $x$ and $y$. As customary, the
south and north poles in axis $j_3$ of the Poincar\'{e} sphere
represent circularly polarized light, while linear polarizations
of different azimuths are distributed along the equator, with
linear polarization along the Cartesian axes $x$ and $y$ located at
the antipodal points of the axis $j_1$ (i.e., $j_1 = \pm j_0$).

\begin{figure}
\begin{center}
\includegraphics[width=6cm]{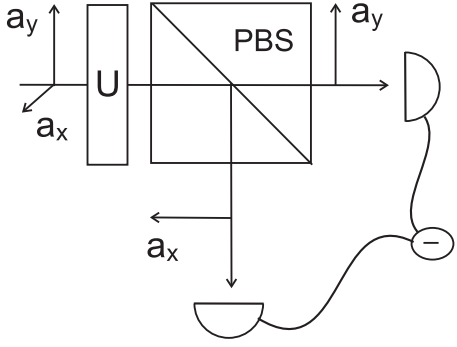}
\end{center}
\caption{Illustration of the scheme for the polarimetric
measurement of the covariance matrix.}
\end{figure}

The most simple polarization measurement is the measurement of
$j_1 = (a^\dagger_x a_x - a^\dagger_y a_y )/2$ as the difference between the
field intensities after a polarizing beam splitter, as illustrated
in Fig. 2. In this case the transformations
(\ref{Ukm}) correspond to phase plates and Faraday rotations
placed before the polarizing beam splitter that transform the
measurement of $j_1$ in the output fields into the measurement
of $j_{k \pm \ell}$ in the input fields. More specifically,
\begin{eqnarray}
\label{br}
& j_{1 \pm 2} = U_{3,\pm 4}^\dagger j_1 U_{3,\pm 4}, &
\nonumber \\
& j_{1 \pm 3} = U_{2,\mp 4}^\dagger j_1 U_{2 ,\mp 4}, &
\nonumber \\
& j_{2 \pm 3} = U_{2, \mp 2}^\dagger U_{3, 4}^\dagger
j_1 U_{3, 4} U_{2, \mp 2} . &
\end{eqnarray}
The transformations $U_{3,\pm 4}$ are Faraday rotations producing
a phase different shift of $\pm \pi/4$ between dextro and levo
circularly polarized modes. This produces a rotation of angle
$\pm \pi/8$ of the azimuth of linearly polarized light, which is
a rotation of the Poincar\'{e} sphere of angle $\pm \pi/4$ along
the north-south axis. On the other hand, the transformations
$U_{2,\pm 2}$ and $U_{2,\pm 4}$ can be implemented by phase plates
introducing phase-difference shifts of $\pm \pi/2$ and $\pm \pi/4$,
respectively, the phase-plate axes forming $\pm \pi/4$ with the
Cartesian axes $x$ and $y$.

\subsection{Two-beam interferometry}

Two-beam interferometry can be embedded in this same framework
by considering that the complex amplitudes $a_{1,2}$ represent
two interfering modes with the same polarization state and
propagating along different directions. In this case, the
simplest measurement is $j_3$, since it represents the
difference of intensities between the two waves $a_{1,2}$.
Otherwise, the same relations (\ref{br}) hold simply by the
cyclic permutation $(1,2,3) \rightarrow (3,1,2)$ for the
indices $k$ and $\ell$ in $j_{k \pm \ell}$, $j_k$, and
$U_{k,\pm m}$.

The transformations (\ref{Ukm}) represent in general lossless
beam splitters and phase shifts. In terms of input-output
relations (\ref{io}) we get for $U_{2, \pm m}$ the following
unitary matrices relating input and output complex amplitudes
\begin{equation}
\label{Ui2}
\mathcal{U}_{2, \pm m} = \pmatrix{\cos (\theta_{\pm} /2) &
\sin (\theta_{\pm} /2) \cr - \sin (\theta_{\pm} /2) &
\cos (\theta_{\pm} /2)} ,
\end{equation}
while for $U_{1, \pm m}$,
\begin{equation}
\label{Ui1}
\mathcal{U}_{1, \pm m} = \pmatrix{\cos (\theta_{\pm} /2) &
i \sin (\theta_{\pm} /2) \cr i \sin (\theta_{\pm} /2) &
\cos (\theta_{\pm} /2)}.
\end{equation}

In Fig. 3 we illustrate how these transformations may be
implemented with very simple elements such as symmetric
beam splitters (SBS) and phase-difference shifts (PDS),
described by the unitary matrices
\begin{eqnarray}
N_\textrm{SBS} &=& \frac{1}{\sqrt{2}} \pmatrix{1 & i \cr
i & 1},\nonumber \\
N_\textrm{PDS} &=&  \pmatrix{\exp{(i \phi)} & 0 \cr 0 &
\exp{(-i \phi)}}.
\end{eqnarray}
For $\mathcal{U}_{2, \pm m}$ we have the parameters
$\varphi = \pi/2$, $\phi = \theta_{\pm m}/2 - \pi/2$,
and $\delta = 0$, while for $\mathcal{U}_{1, \pm m}$ the
parameters are $\varphi = \delta = \pi/4$ and the same
$\phi = \theta_{\pm m}/2 - \pi/2$.

\begin{figure}
\begin{center}
\includegraphics[width=6cm]{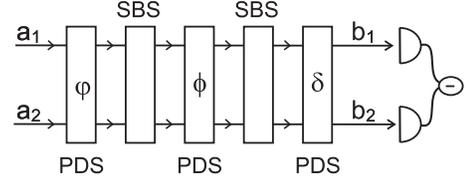}
\end{center}
\caption{Illustration of the interferometric realization of
the transformations (\ref{Ui2}) and (\ref{Ui1}) using
symmetric beam splitters (SBS) and phase-difference shifts
(PDS) exclusively.}
\end{figure}

\subsection{Two-level atoms}

In this case the physical situation corresponds to a collection
of $N$ two-level atoms (with ground $| g \rangle$ and excited
$| e \rangle$ levels) interacting with a classical field
$\bm{E}=\bm{E_0}\cos(\omega t)$. By assuming that the
coupling between atoms can be neglected the total Hamiltonian
is given by the sum of individual Hamiltonians (in units $\hbar =1$
for simplicity),
\begin{equation}
\label{Htotal}
H = \sum_{k=1}^N h_k ,
\end{equation}
being
\begin{equation}
\label{hk}
h_k = \frac{\omega_0}{2}\sigma_3^{(k)} - \Omega \left [
\sigma_{-}^{(k)}+\sigma_{+}^{(k)}\right]\cos(\omega t) ,
\end{equation}
where, for each atom $k$,
\begin{equation}
\sigma_3 = | e \rangle \langle e | - | g \rangle \langle g |,
\quad
\sigma_- = \sigma_+^\dagger = | g \rangle \langle e |
\end{equation}
are the corresponding Pauli and ladder matrices with $\sigma_\pm
= (\sigma_1 \pm i \sigma_2 )/2$. The first term in Eq. (\ref{hk})
is the free-evolution Hamiltonian for each atom and the second
one is the coupling with the classical field at dipolar
approximation. The Rabi frequency $\Omega = \langle g|\hat{\bm{d}}
|e\rangle\cdot \bm{E_0}$ has been assumed to be real. On the
regimen $\omega \simeq \omega_0 \gg \Omega$ is usually used to consider
the rotating wave approximation by neglecting the counter-rotating
terms $\sigma_+\exp(i\omega t)$ and $\sigma_- \exp(-i\omega t)$
in the above Hamiltonians \cite{SZ}
\begin{eqnarray}
\label{hkr}
h_k \simeq &\frac{\omega_0}{2}&\sigma_3^{(k)} - \frac{\Omega}{2}
\left [ \sigma_{-}^{(k)}\exp(i\omega t)+\sigma_{+}^{(k)}
\exp(-i\omega t) \right ] \nonumber\\
=&\frac{\omega_0}{2}&\sigma_3^{(k)} -
\frac{\Omega}{2} \left [ \sigma_1^{(k)}\cos(\omega t) +
\sigma_2^{(k)} \sin(\omega t)\right ].
\end{eqnarray}
From Eqs. (\ref{Htotal}) and (\ref{hkr}) the total Hamiltonian
can be written in terms of the total angular momentum $\bm{j}=
\sum_k \bm{\sigma}^{(k)}/2$ as
\begin{equation}
H = \omega_0 j_3 - \Omega \left [ j_1 \cos(\omega t) + j_2 \sin (
\omega t) \right ] = U_3^\dagger \left [ \omega_0 j_3 - \Omega j_1
\right ] U_3 ,
\end{equation}
where $U_3=\exp(i\omega tj_3)$ is a rotation around the $j_3$ axis.

It is customary to change the picture by the unitary transform
$U_3 = \exp(i \omega tj_3)$ in order to remove the time dependence
of the Hamiltonian
\begin{equation}
|\psi(t)\rangle\rightarrow|\tilde{\psi}(t)\rangle =
U_3 | \psi(t) \rangle ,
\end{equation}
with $|\tilde{\psi}(0)\rangle=|\psi(0)\rangle$. The time-evolution
equation in this picture is
\begin{equation}
i\frac{d}{dt}|\tilde{\psi}(t)\rangle=\left[(\omega_0-\omega)
j_3-\Omega j_1 \right ] |\tilde{\psi}(t)\rangle ,
\end{equation}
so the operator $U(t_2, t_1)$ performing the time evolution
between $t_1 =0$ and $t_2 =t$ in the Schr\"{o}dinger picture is
\begin{equation}
U(t,0)=\exp(-i\omega tj_3)\exp \{-i \left [(\omega_0-\omega)
j_3-\Omega j_1 \right ] t \} .
\end{equation}
This operator reduces to a very simple product of SU(2)
transformations when the radiation field is in resonance
$\omega = \omega_0$:
\begin{equation}
U_{\mathrm{res}}(t,0) = \exp(-i\omega_0 tj_3)
\exp(i\Omega tj_1) .
\end{equation}
On the other hand, if the external field is switched off $(\Omega=0)$,
the free evolution is given by
\begin{equation}
U_{\mathrm{free}}(t)=\exp(-i\omega_0 tj_3).
\end{equation}
Therefore, in this case the transformations $U_{k \pm m}$ are
obtained by combining time intervals of resonance pulses and
free evolution, as is used in Ramsey spectroscopy \cite{Rm}.

As in the interferometric case above, the simplest measurement
is $j_3$ again, this is the difference of populations between
the two levels of the atoms (nevertheless, see Ref. \cite{asc}
for other light-mediated atomic-spin measuring schemes).
More explicitly $U_{2,\pm m}$ can be achieved as
\begin{equation}
U_{2,\pm m}=U_{\mathrm{free}}(t_{\pi/2}-t_{\pm m})
U_{\mathrm{res}} (t_{\pm m},0)U_{\mathrm{free}}(t_{-\pi/2}) ,
\end{equation}
with $\omega_0 t_{\pm \pi/2} = \pm \pi/2 \; \mathrm{mod}
( 2 \pi)$ and, in order to deal always with positive time
intervals, $\Omega t_{m}=\pi/m$ and $\Omega t_{-m}=(2m-1)\pi/m$.
Similarly $U_{1,\pm m}$ can be achieved as
\begin{equation}
U_{1,\pm m}=U_{\mathrm{free}}(t_{2 \pi }-t_{\pm m})
U_{\mathrm{res}}(t_{\pm m},0) ,
\end{equation}
with $\omega_0 t_{2\pi} = 2\pi \; \mathrm{mod} ( 2 \pi)$ and
the same
$t_{\pm m}$ above. We stress that the $\mathrm{mod} ( 2 \pi)$
freedom should be used to obtain always positive time intervals.

In these equations the following relation is useful
\begin{equation}
\exp( i \varphi j_2 ) = \exp \left ( -i\frac{\pi}{2}j_3 \right )
\exp( i \varphi j_1 ) \exp \left ( i \frac{\pi}{2} j_3 \right ) ,
\end{equation}
which can be derived from the relations in Sec. IIC for $\theta
= \pi/2$ and $V= \sigma_3 /2$, so that the rotation matrix in
Eqs. (\ref{rot}) and (\ref{Rkl}) becomes
\begin{equation}
R = \pmatrix{0 & 1 & 0 \cr - 1 & 0 & 0 \cr 0 & 0 & 1} .
\end{equation}

\section{Simultaneous measurements}

In this section we show that an interferometric noisy simultaneous
measurement of the three components of $\bm{j}$ provides a simple
and exact determination of the covariance matrix with a single
apparatus (similar schemes may be developed for the other contexts). The subject of simultaneous
measurements of noncommuting observables has a long history,
being closely related to basic issues of quantum theory
such as generalized measurements, state reconstruction,
complementarity, uncertainty relations, etc. \cite{GM}.
In our case we consider the 12-port scheme illustrated in Fig. 4
\cite{YP,RI00}. The two input signal modes are $a_1$ and $a_2$, while
the input modes $a_{10}$, $a_{20}$, $a^\prime_{10}$ and $a^\prime_{20}$
are auxiliary modes always in the vacuum state.

\begin{figure}
\begin{center}
\includegraphics[width=8cm]{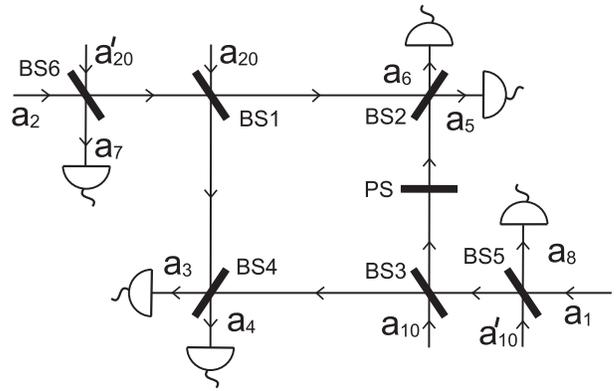}
\end{center}
\caption{Illustration of the 12-port scheme.}
\end{figure}

For definiteness and to simplify formulas let us consider that
beam splitters BS1, BS2, BS3 and BS4 are 50\% with real transmission
and reflection coefficients and a $\pi$ phase change in the upper
side reflections. BS5 and BS6 are identical with real transmission
$t$ and reflection $r$ coefficients, with $t \neq r$, and a $\pi$
phase shift in the upper side reflections. Finally PS is a $\pi/2$
phase shift. The relation between the input and output complex
amplitudes is \cite{YP}
\begin{eqnarray}
a_3 & = & \frac{1}{2} \left ( t a_1  - t a_2 + a_{10} - a_{20} +
r a^\prime_{10} + r a^\prime_{20} \right ) , \nonumber \\
a_4 & = & \frac{1}{2} \left ( t a_1  + t a_2 + a_{10} + a_{20} +
r a^\prime_{10} - r a^\prime_{20} \right ) , \nonumber \\
a_5 & = & \frac{1}{2} \left ( -it a_1  + t a_2 + i a_{10} -
a_{20} - i r a^\prime_{10} - r a^\prime_{20} \right ) ,\nonumber \\
a_6 & = & \frac{1}{2} \left ( -it a_1  - t a_2 + i a_{10} + a_{20}
- i r a^\prime_{10} + r a^\prime_{20} \right ) , \nonumber \\
a_7 & = & r a_2 + t a^\prime_{20} , \nonumber \\
a_8 & = & - r a_1 + t a^\prime_{10} .
\end{eqnarray}

In the classical domain the vacuum state implies that $a_{10} =
a_{20} = a^\prime_{10} = a^\prime_{20} = 0$, so we have the
noiseless simultaneous measurement of all the components
(\ref{So}) via the detection of the six output intensities
$I_j = a^\ast_j a_j$, $j=3,\ldots,8$, in the form
\begin{eqnarray}
j_0 = \frac{1}{2} \sum_{j=3}^8 I_j , & & j_1 = \frac{1}{t^2}
\left ( I_4 - I_3 \right ), \nonumber \\
j_2 = \frac{1}{t^2} \left ( I_6 - I_5 \right ) , & &
j_3 = \frac{1}{2 r^2} \left ( I_8 - I_7 \right ) .
\end{eqnarray}
With this we can compute the whole covariance matrix $M$ by
determining the variances and correlations between the
output intensities $I_j$.

In the quantum case the amplitudes of the auxiliary modes
$a_{10}$, $a_{20}$, $a^\prime_{10}$ and $a^\prime_{20}$
cannot be taken as zero since the complex amplitude of the
vacuum fluctuates. In other words, simultaneous exact
measurements of noncommuting operators are forbidden by
commutation relations. Nevertheless, it is still possible
to extract useful and reliable information from simultaneous
noisy measurements. To this end let us define the commuting
measured observables
\begin{eqnarray}
& \tilde{j}_1 = \frac{1}{t^2} \left ( a^\dagger_4 a_4 -
a^\dagger_3 a_3 \right ), \nonumber \\
& \tilde{j}_2 = \frac{1}{t^2} \left ( a^\dagger_6 a_6
- a^\dagger_5 a_5 \right ), & \nonumber \\
& \tilde{j}_3 = \frac{1}{2 r^2} \left ( a^\dagger_8 a_8 -
a^\dagger_7 a_7 \right ) , &
\end{eqnarray}
as providing a noisy joint measurement of the operators
(\ref{So}). We do not include $j_0$ because this measurement
is actually exact and noiseless because of conservation of
total photon number between the input and output (the
auxiliary modes are in an eigenstate of the number operator).

In Ref. \cite{YP} it was shown that for the mean values and
variances we have
\begin{equation}
\langle \tilde{\bm{j}} \rangle = \langle \bm{j} \rangle 
\end{equation}
and
\begin{eqnarray}
\label{dt}
\left ( \Delta \tilde{j}_1 \right )^2 & = &
\left ( \Delta j_1 \right )^2 + \frac{1+r^2}{2 t^2}
\langle j_0 \rangle , \nonumber \\
\left ( \Delta \tilde{j}_2 \right )^2 & = &
\left ( \Delta j_2 \right )^2 + \frac{1+r^2}{2 t^2}
\langle j_0 \rangle , \nonumber \\
\left ( \Delta \tilde{j}_3 \right )^2 & = &
\left ( \Delta j_3 \right )^2 + \frac{t^2}{2 r^2}
\langle j_0 \rangle ,
\end{eqnarray}
so that the diagonal terms of the covariance matrix
$\Delta j_{1,2,3}$ can be determined simply and exactly
from $\Delta \tilde{j}_{1,2,3}$ and $\langle j_0 \rangle$.
Concerning the nondiagonal terms, it can be seen that the
following exact relations hold for all $k \neq \ell$:
\begin{equation}
\label{ndt}
\langle \tilde{j}_\ell \tilde{j}_k \rangle =
\langle \tilde{j}_k \tilde{j}_\ell \rangle =
\langle : j_\ell j_k : \rangle =
\frac{1}{2} \langle \left ( j_k j_\ell + j_\ell j_k \right )
\rangle ,
\end{equation}
where $: \; :$ denotes normal ordering. To derive this last
relation it can be taken into account that $a^\dagger_k a_k
a^\dagger_\ell a_\ell = a^\dagger_k a^\dagger_\ell  a_k
a_\ell$ for $k \neq \ell$ in order to express $\tilde{j}_\ell
\tilde{j}_k $ in normal order. This is useful since this
automatically removes the operators of the auxiliary modes
in the vacuum state, leading to $\langle \tilde{j}_\ell
\tilde{j}_k \rangle = \langle : j_\ell j_k : \rangle$. Then,
the last equality in Eq. (\ref{ndt}) can be proved by direct
computation.

Therefore we get that the statistics of $\tilde{\bm{j}}$
allow one to determine the exact mean values and the covariance
matrix for $\bm{j}$. The variances of $\tilde{\bm{j}}$
present an excess of fluctuations caused by the vacuum in
the auxiliary modes that can be easily subtracted or
compensated. This is particularly simple for $t=\sqrt{2/3}$,
$r=\sqrt{1/3}$, since in such a case
\begin{equation}
M = \tilde{M} - \langle j_0 \rangle \mathcal{I},
\end{equation}
where $M$ and $\tilde{M}$ are the correlation matrices for
the $\bm{j}$ and $\tilde{\bm{j}}$ operators, respectively,
and $\mathcal{I}$ is the $3 \times 3$ identity matrix.

It is worth stressing the simplicity of this method since it
provides complete information via the measurement of just four
observables ($\tilde{\bm{j}}$, $j_0$) instead of the six
observables of the general method in Eq. (\ref{mc}). Moreover,
the four observables ($\tilde{\bm{j}}$, $j_0$) are measured in
a single experimental arrangement.

The above relations (\ref{ndt}) can be regarded as a
correspondence between classical variables (the outputs
of measuring $\tilde{\bm{j}}$) and quantum mechanical
operators $\bm{j}$. In particular, Eq. (\ref{ndt}) is
actually an angular momentum version of the Wigner
\cite{W} and Terlesky-Margenau-Hill \cite{TMH}
correspondences between products of classical variables
and symmetric operator orderings.

Similar schemes may be developed for the polarimetric context.
The same interferometric scheme above is valid if the signal
modes $a_{1,2}$ are the polarization components $a_{x,y}$.
A more polarimetric scheme allowing the noisy simultaneous
measurement of $j_1$, $j_2$, and $j_3$ is outlined in Fig. 5.
The two beam splitters BS provide three copies of the original
beam by mixing with the vacuum, which are directed to three
detectors $D$ that are essentially of the form of the $j_1$
measuring scheme in Fig. 2. The transformations $U_{2,2}$ and
$U_{3,2}$ placed in front of them transform the measurement
of $j_1$ into measurements of $j_3$ and $j_2$.

\begin{figure}
\begin{center}
\includegraphics[width=5cm]{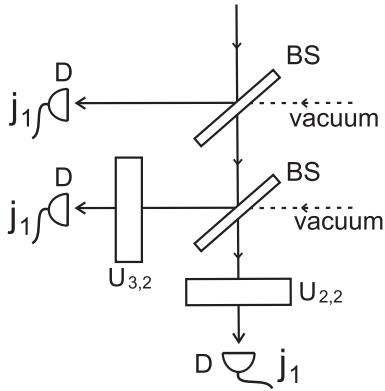}
\end{center}
\caption{Illustration of a practical scheme for the noisy
simultaneous measurement of $j_1$, $j_2$, and $j_3$ in a
polarimetric context.}
\end{figure}

\section{Bright limit}

Focusing on the bosonic realization, when the state of one
of the modes is known, the above measuring schemes provide
information about the statistical properties of the other
mode. Let us examine this issue by considering for
definiteness that the system state factorizes $\rho = \rho_1
\otimes \rho_2$, being $\rho_1 =  | \alpha \rangle \langle
\alpha |$, where $| \alpha \rangle$ is a coherent state
$a_1 | \alpha \rangle = \alpha | \alpha \rangle$ with real
$\alpha$ for simplicity. In such a case we have
\begin{equation}
\langle j_1 \rangle = \alpha  \langle X \rangle ,
\quad
\langle j_2 \rangle = \alpha  \langle Y \rangle ,
\quad
\langle j_3 \rangle = \frac{1}{2} \left ( \alpha^2
- \langle n \rangle \right ) ,
\end{equation}
where $X$ and $Y$ are the quadrature operators of mode $a_2$,
\begin{equation}
X = \frac{1}{2} \left ( a^\dagger_2 + a_2 \right ) ,
\quad
Y = \frac{i}{2} \left ( a^\dagger_2 - a_2 \right ) ,
\end{equation}
and $n = a^\dagger_2 a_2$ is the number operator. Concerning
the covariance matrix we have the following exact series in
powers of $\alpha$, valid for any $\alpha$,
\begin{equation}
\label{Ma}
M = \alpha^2 M_2 + \alpha M_1 + M_0 ,
\end{equation}
with
\begin{equation}
\label{M2}{\scriptsize
M_2 = \pmatrix{\left ( \Delta X \right )^2 & \frac{1}{2}
\langle \left ( XY + YX \right ) \rangle - \langle X \rangle
\langle Y \rangle & 0 \cr \frac{1}{2} \langle  \left ( XY +
YX \right ) \rangle - \langle X \rangle
\langle Y \rangle & \left ( \Delta Y \right )^2  & 0 \cr 0 &
0 & \frac{1}{4}}} ,
\end{equation}
\begin{equation}
M_1 = \frac{1}{4} \pmatrix{0 & 0 & f(X) \cr 0 & 0 & f(Y) \cr
f(X) & f(Y) & 0} ,
\end{equation}
with
$f(A) = \langle A \rangle + 2 \langle n \rangle \langle A
\rangle - \langle \left ( A n + n A \right ) \rangle$ and
\begin{equation}
M_0 = \frac{1}{4} \pmatrix{\langle n \rangle & 0 & 0 \cr
0 & \langle n \rangle & 0 \cr 0 & 0 & \left ( \Delta n
\right )^2} .
\end{equation}

In the bright limit $\alpha \rightarrow \infty$ the
$\alpha$-leading term in Eq. (\ref{Ma}) is $M_2$ so that
the angular-momentum covariance matrix becomes essentially
the quadrature covariance matrix of mode $a_2$. This is
because in such a limit $X$ and $Y$ become the Cartesian
coordinates of the plane tangent to the Poincar\'{e} sphere
at point $\langle \bm{j} \rangle / \langle j_0 \rangle
\simeq (0,0,1)$ \cite{LK}. In the same conditions we can
consider the bright limit of the 12-port detection
scheme in Sec. IV, where mode $a_1$ plays the role of the
local oscillator. When the local oscillator is in a coherent
state $| \alpha \rangle$ with $\alpha \rightarrow \infty$
we have
\begin{eqnarray}
& \left ( \Delta \tilde{j}_1 \right )^2 \simeq \alpha^2
\left [ \left ( \Delta X \right )^2 + \frac{1+r^2}{4 t^2}
\right ] ,& \nonumber \\
& \left ( \Delta \tilde{j}_2 \right )^2 \simeq \alpha^2
\left [ \left ( \Delta Y  \right )^2 + \frac{1+r^2}{4 t^2}
\right ] ,& \nonumber \\
& \langle \tilde{j_1} \tilde{j_2} \rangle =
\langle \tilde{j_2} \tilde{j_1} \rangle = \frac{\alpha^2}{2}
\langle \left ( X Y + Y X \right ) \rangle . &
\end{eqnarray}
This agrees well with Eq. (\ref{M2}), as well as with
previous analyses of the bright limit of multi-port homodyne
detection \cite{BL}. Analyses of homodyne detection with
finite local oscillator are also available \cite{YA}.

Finally we point out that these results are valid in the
general case beyond two-mode bosonic realizations. This
holds via the idea of group contraction that applies when
the state of the system remains in a small enough region
of the Poincar\'{e} sphere that can be well approximated
by the tangent plane \cite{CS}.

\section{Conclusions}

We have provided some general schemes for the practical
determination of angular-momentum covariance matrices in
different contexts. They allow us to determine a global and
SU(2)-invariant characterization of angular-momentum
fluctuations via principal components. This can be of
interest, for example, for unambiguous, reference-free
characterization of SU(2) squeezing with applications in
quantum metrology, detection of many-body and continuous-variable
entanglement, and light-mediated detection of atomic-spin
correlations.

In particular, we have shown that this task can be accomplished
in an exact and simple way by noisy simultaneous measurement
of angular-momentum components. This provides complete
information via the measurement of just four observables,
instead of the six observables required by the general method.
Moreover, they are measured in a single
experimental arrangement.

It is worth stressing the simplicity of this scheme. The
minimum number of measured observables required to determine
the principal variances is three provided we know in advance
the principal components. Just by adding a fourth measurement
we no longer need to know the principal components, since we
can gather enough information to determine the whole covariance
matrix. This includes at once complete information about principal components,
principal variances, and the variances of angular-momentum
components along arbitrary directions.

\section*{Acknowledgments}

A.R. acknowledges financial support from the University
of Hertfordshire and the EU Integrated Project QAP. A.L.
acknowledges support from Project No. FIS2008-01267/FIS
of the Spanish Dirección General de Investigación del Ministerio
de Ciencia e Innovación.

\end{document}